# Low Energy Cosmic Rays and the Disturbed Magnetosphere

### K. Kudela, R. Bucik

*Institute of Experimental Physics, Slovak Academy of Sciences, Kosice, Slovakia, kkudela@upjs.sk*

*Low energy galactic cosmic rays as well as particles accelerated to high energies either at the solar surface or in the interplanetary medium have access to the atmosphere above a given position on the Earth depending upon the state of the magnetosphere. The interpretation of the cosmic ray anisotropy, deduced from the neutron monitor (NM) network, must assume the variability of the magnetospheric configuration. Along with a short review of changes of the geomagnetic cutoffs in the disturbed magnetosphere reported in the earlier papers, we present the results of computations of transmissivity function and asymptotic directions for selected points on the ground and for a low altitude polar orbiting satellite as well. The computations, based on different available models of geomagnetic field of external sources are performed for quiet time periods and for strong geomagnetic disturbances occurred in 2003 and 2004.*

## Introduction

Access of low energy cosmic rays to the Earth is controlled by the geomagnetic field. This effect is usually estimated by particle trajectory tracing from the point of observation in the model geomagnetic field (e.g. [1]). The computational results are compared with the measurements both on ground (e.g. [2-4] among others) as well as on low altitude satellites (e.g. [5-7]). During the enhanced geomagnetic activity, changes of cutoffs and asymptotic directions occur (e.g. [8]). Relevance of cosmic rays as one of the parameters for the forecasts of space weather effects is checked during the past decade and special networks/systems are created (e.g. [9-17]). Also the direct relations of cosmic rays to space weather are studied (e.g. [18-21]). The interpretation of cosmic ray variability and/or anisotropy requires good estimates of the transmissivity and asymptotic characteristics of cosmic rays.

Intervals of October, November 2003 & November 2004 with strong geomagnetic disturbances are useful for checking the changes of magnetospheric transmissivity for galactic and solar cosmic rays. For this purpose, in addition to the cosmic ray obtained from the NMs, here we use measurements on Russian low altitude polar orbiting satellite CORONAS-F by a detector with large geometrical factor sufficient to observe variations of low energy galactic cosmic rays [22,23]. The SKL complex of instruments is also used for mapping the boundary of solar electron penetration to high latitudes [24].

Most of the papers published earlier on cutoff rigidity computations and/or geomagnetic transmissivity changes, are based on the "static" geomagnetic field models combining the geomagnetic field of internal and external sources. A new model, suggested for the periods of strong geomagnetic storms was published by Tsyganenko and Sitnov [25]. Its difference from the earlier models is the principle sources of the external magnetic field are driven by a separate variables, derived from the "prehistory" of the disturbance, i.e. as a time integral of a combination of geo-effective parameters as solar wind speed, density and north-south component of Interplanetary Magnetic Field (IMF).

We present computations of the geomagnetic transmissivity for three selected periods with different disturbances of the magnetosphere, and we report the increases of solar cosmic rays and low energy galactic particles observed on the ground and on the satellite due to the improved transmissivity of magnetosphere. In addition to the models used previously, the computations here are done also with a new model [25] depending on the availability of input data.

## Method

We use a method, similar to earlier ones (e.g. [1]), tracing the cosmic ray particle trajectory from a given point on the Earth's surface with the reversed charge sign and velocity vector and numerically solving the equation of motion in the model field **B**. The details of the computation method and dependence of the result on parameters of computations are described in [26-28].

For **B** we use the modified 'Geopack 2003' subroutines (http://nssdc.gsfc.nasa.gov/space/model/magnetos/data-based/modeling.html) based on International Geomagnetic Reference Field (IGRF) model (http://www.ngdc.noaa.gov/IAGA/vmod/igrf.html). The IGRF coefficients having order/degree up to 10 are used. To look into the comparison with cosmic ray intensity variations we have used the computed values of the upper and effective cutoff rigidities as characteristics of magnetospheric transmissivity. Their definitions are given in [29]. When the first of the values depicts the highest rigidity transition between the forbidden and allowed trajectory, the second one depends on the structure of penumbra. The computations are done for vertical direction of cosmic ray access and the step in rigidity is 0.001 GV.

We have performed the computations of the upper and effective cutoffs using three approaches- (1) Tsyganenko Ts89 model [30], (2) extension of Ts89 with Dst [31] and (3) the new Tsyganenko Ts04 model [25]. As the first two models are parameterized by Kp and by instant Dst value, the inputs to the second model are derived from prehistory of solar wind and IMF parameters.

## Interval 1 (October 28 - November 1, 2003)

The cosmic ray (CR hereafter) intensity variations as measured by low latitude NMs at Haleakala and Mexico are shown in Fig. 1. Increases in the intensity with maximum at

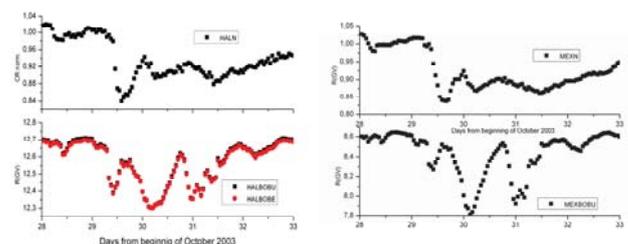

*Fig.1.* Upper panels: *Haleakala (HALN) and Mexico (MEXN) NM normalized CR intensity. Lower panels: vertical cutoff rigidity (upper-U and effective-E by Ts89+Dst model field) variations during interval 1.*

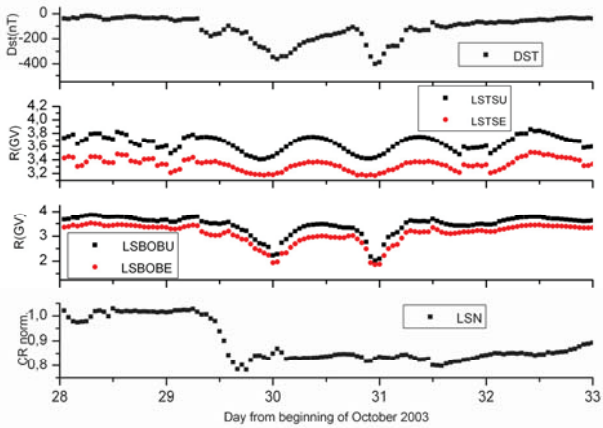

*Fig.2. From top to bottom: Dst; upper-U and effective-E vertical cutoff rigidities by Ts89 (LSTS) and Ts89+Dst model (LSBOB); Lomnicky Stit (LSN) CR intensity, normalized to 00-12 UT of Oct.28, 2003.*

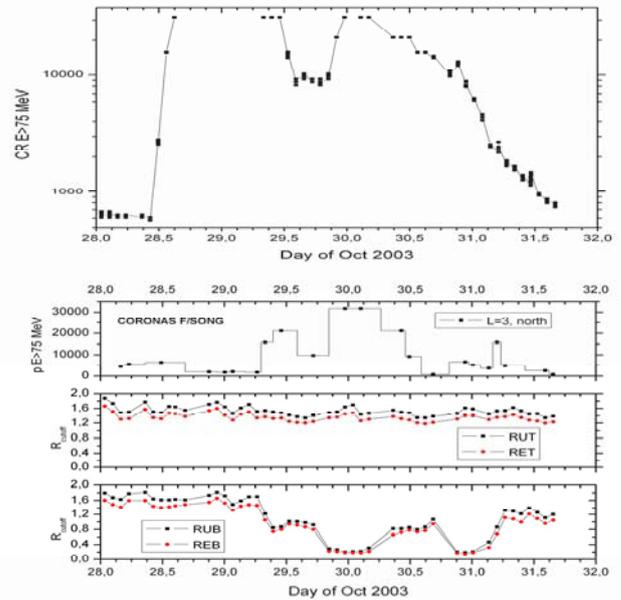

*Fig.3. The event on October 28-31, 2003 seen at high latitudes (L>15 selections, upper panel) and at L=3 in the northern hemisphere (2nd panel from top) by CORONAS-F satellite. The two next lower panels show the computed vertical cutoffs for Ts89 model (RUT - upper, RET - effective) and for Ts89+Dst (RUB - upper, REB - effective) respectively.*

~01 UT on October 30 are observed during the decreases of both the Dst (after 00 UT on Oct. 30) and computed by Ts89+Dst external field model vertical cutoff rigidities, upper and effective. An example for Lomnicky Stit NM is given in Fig. 2. The vertical rigidities with model Ts89 shows local time variations. The CR intensity enhancements during Dst decreases and during reductions of computed cutoffs using model Ts89+Dst are not well pronounced here.

Low altitude (~500 km), polar orbiting satellite CORONAS-F, launched on July 31, 2001 and finished its mission on December 4, 2005 contained SONG instrument, dedicated for measuring energetic neutral radiation from the Sun [22]. Along with that, the scintillator detector provides the measurements of protons with energy >75 MeV and electrons >40 MeV. It has a large geometrical factor (2000 cm$^2$sr, channel 1 in the following). This allows to measure on the orbit the variations of low energy cosmic rays especially during the solar flare events and also during the Forbush decreases. SONG on CORONAS-F has one more high energy proton channel, namely, for protons Ep=200-300 MeV (channel 2). However, at an earlier experiment on CORONAS-I we showed that the decreases of galactic cosmic rays could have been observed by means of the similar instrument at the channel with even higher energies [23].

Fig. 3 shows the time evolution of proton measurements by channel 1 of SONG instrument during the interval of October 28-31, 2003. At high latitudes (north, L>15) the large SEP event is observed. Penetration of protons E> 75 MeV to L=3 is related to local cutoff rigidity decrease computed for CORONAS-F orbit by Ts89+Dst model. Model Ts89 is not sufficient to explain the variations of the protons penetration. Computations using the Dst extension of that model show two minima in cutoff time profile which seem to be approximate but not exactly corresponding to the measurement.

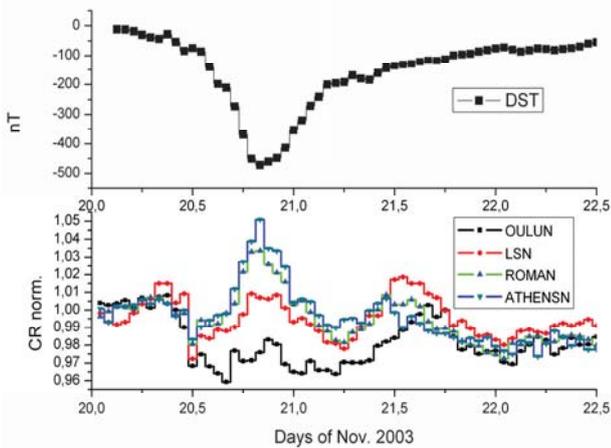
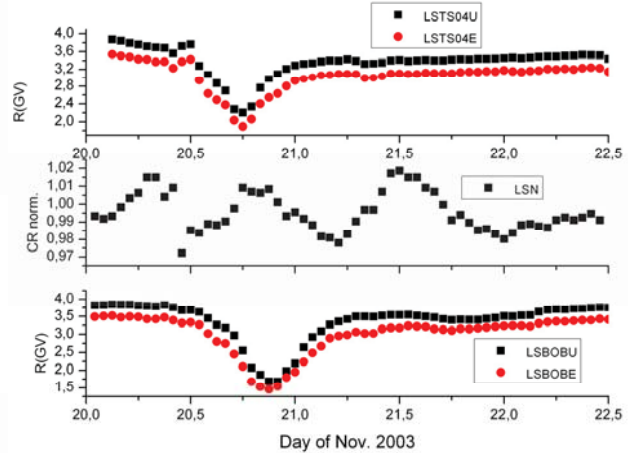

*Fig.4. Left: Dst –upper panel; Neutron monitor time profiles at 4 European Observatories (Oulu - OULUN, Lomnicky Stit - LSN, Rome - ROMAN, Athens - ATHENSN) during the Dst decrease on November 20-21, 2003, normalized to the mean 00-12 UT on November 20. Right: Lomnicky Stit NM normalized counting rate profile (middle panel) compared with the effective and upper cut-offs obtained from trajectory computations. The vertical cutoffs are computed for the approach (2) (LSBOBU – upper, LSBOBE – effective) and for approach (3) (LSTS04U – upper, LSTS04E – effective) of geomagnetic field model.*

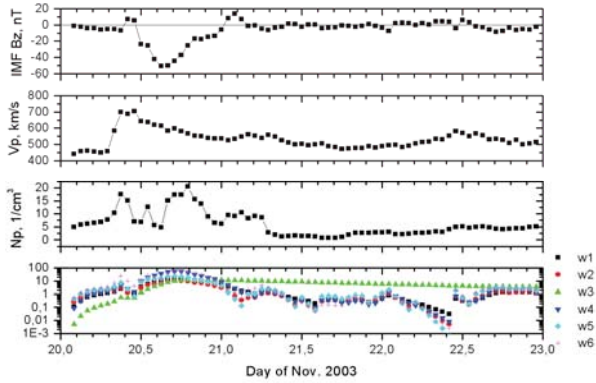

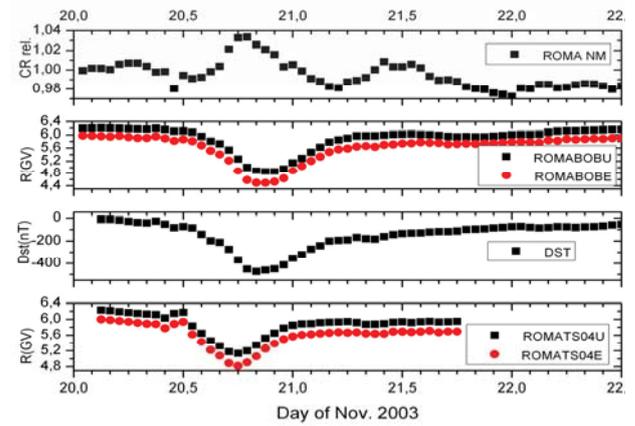

*Fig.5. Input parameters W1(inner tail current), W2(outer tail current), W3(symmetrical ring current), W4(partial ring current), W5(field-aligned current-region 1), W6(field-aligned current-region 2) for Ts04 model - constructed from the time profiles of solar wind densit Np, speed Vp, IMF Bz by Tsyganenko and Sitnov [25] formula (7) modified for one – hour data.*

*Fig.6. Rome NM normalized counting rate profile compared with the effective and upper cut-offs obtained from trajectory computations. The vertical cutoffs are computed for the approach (2) (ROMABOBU – upper, ROMABOBE – effective) and for approach (3) (ROMATS04U – upper, ROMATS04E – effective) of geomagnetic field model.*

**Interval 2 (November 20-23, 2003)**

The largest decrease of Dst during the years of 2002-2005 was observed on November 20, 2003 (Dst=-471nT). This disturbance affected strongly the transmissivity of the magnetosphere and, similarly to other events [e.g. 2,32] led to the increase of cosmic ray intensity at NMs with relatively high nominal cutoff rigidity as shown in Fig. 4 (left).

For interval 2 the data of solar wind and IMF (see Fig. 5) are available (from http://nssdc.gsfc.nasa.gov/omniweb/). Also model Ts04 is used for cutoff rigidity variations estimate. Fig. 4 (right) demonstrates that for Ts04 model the vertical cutoff reductions are smaller than for Ts89+Dst extension and minima of computed cutoff rigidities for approach (3) rather precede the increase in NM cosmic ray intensity than it is in approach (2). Similar pattern with the better timing of minimum cutoff to the corresponding peak at NM Rome for Ts04 model than for Ts89+Dst extension is shown in Fig. 6.

The geomagnetic transmission for cosmic rays access to the Earth can be quantitatively described by the transmissivity function approach. The transmissivity function (TF) here is defined as the probability that a particle having rigidity (R, R+dR) will access the detector from zenith direction (vertical access). For dR we use 0.1 GV. Fig. 7 displays transmissivity function computed by different external field models for Lomnicky Stit in quite and disturbed times. One can see that shift of TF to lower rigidities during disturbed period varies significantly with different field models. Fig. 8 shows the corresponding structure of asymptotic directions out of penumbra region. The asymptotics are shifted significantly westward, which is particularly well pronounced for the low rigidity part of the arriving particles, and their range is narrowed when the magnetic activity is increased. However, presented asymptotic directions do not show net variations on magnetic activity; variations on local time have been recently reported in [28]. Smooth structure on top panels correspond to the entry in day-side sector, while slightly perturbed one on lower panels is related to the night side entries.

**Interval 3 (November 7-12, 2004)**

This is another strongly disturbed interval when the inputs into Ts04 model are found available. There is observed rather complex, long lasting decrease in cosmic rays shown in Fig. 9 for one middle latitude station. The recovery of CR intensity observed on the ground to the pre-storm level was reached close to the end of November (at Lomnicky Stit). Cosmic ray intensity below the atmospheric threshold (channel 2 at SONG /CORONAS-F) is still depressed as indicated in Fig. 9 low panel.

During the geomagnetic disturbance on November 8, 2004 the transmissivity of magnetosphere increased - one signature of them is seen at Lomnicky Stit while similarly to the second interval analyzed here (November 20-23, 2003) it was better pronounced again at Rome and at Athens NM (not shown here). There was a moderate increase of proton flux with the maximum around the midnight 07/08 November observed at GOES. The Dst was decreasing to -373 nT at 07 UT on November 8. During the Dst decrease the channel 1 of SONG indicated increases when CORONAS-F crossed L=3. Even at the crossing L=2 the increase was observed (not displayed here). The profiles at the crossings of L=3 and the cutoff expectations on CORONAS-F orbit are presented in Fig. 10. Although the increase of counting rate corresponds to the time interval when cutoffs are expected to be <1 GV, the time of maximum does not match exactly the minimum of cutoffs. Contrary to the event on November 20-23, 2003, as changes of both cutoffs were much stronger for Ts89+Dst approach than for Ts04 one, here the differences are not significant.

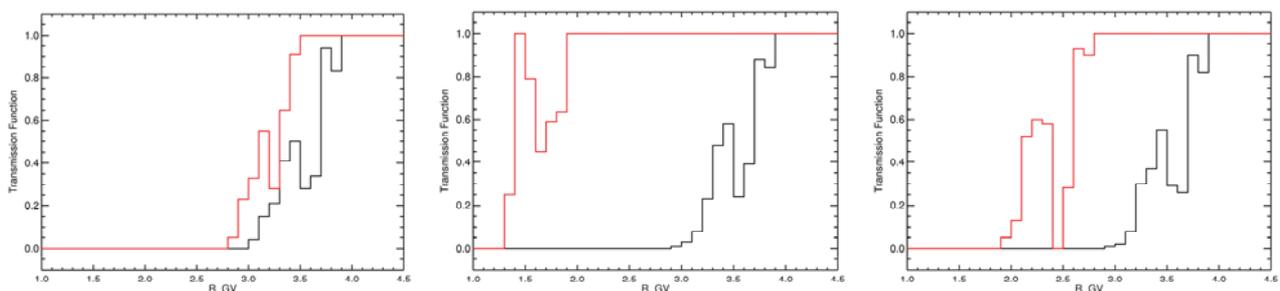

*Fig.7. Transmission functions (for vertical direction) for Lomnicky Stit before the onset of the storm (Nov. 20, 2003, 02 UT, black) and during the Dst minimum (19 UT, red) for three models (from left to right: Ts89, Ts89+Dst, Ts04).*

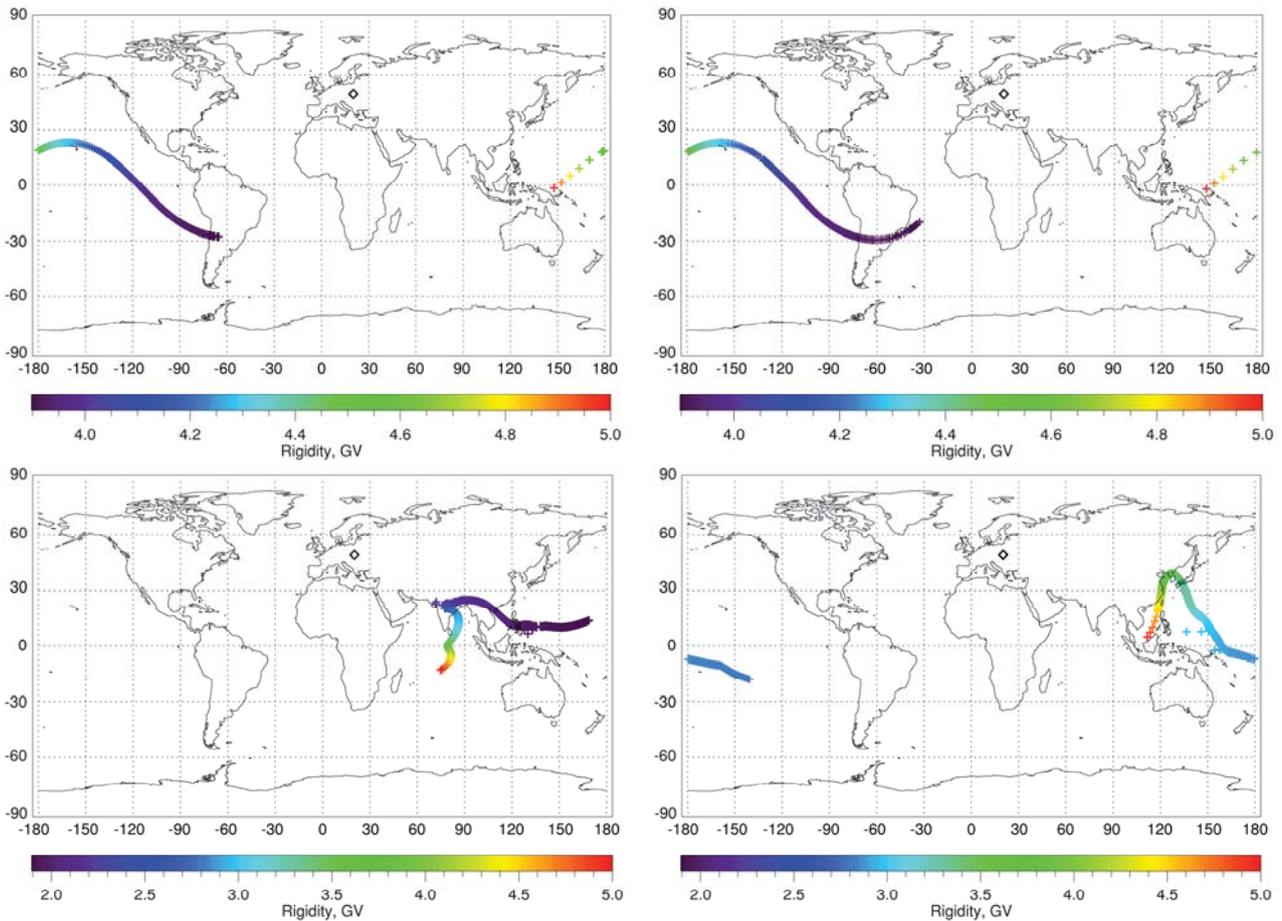

*Fig.8. Asymptotic directions for vertical incident particles in rigidity range between upper rigidity and 5 GV for Lomnicky Stit (black diamond on map) before the disturbance (upper panels, Nov. 20, 2003, 02 UT) and at minimum Dst (lower panels, Nov. 20, 2003, 19 UT) for Ts89+Dst model (left) and Ts04 model (right).*

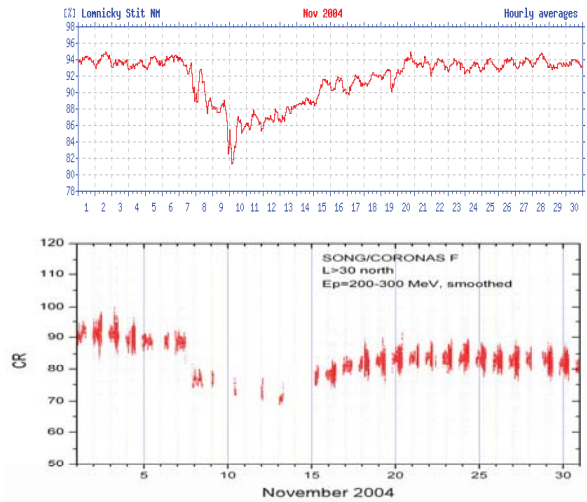

*Fig.9. Cosmic ray decrease observed at Lomnicky Stit neutron monitor (top) and at the high latitude passes on CORONAS-F (low), channel 2 of SONG instrument, during the interplanetary and geomagnetic disturbances in November 2004.*

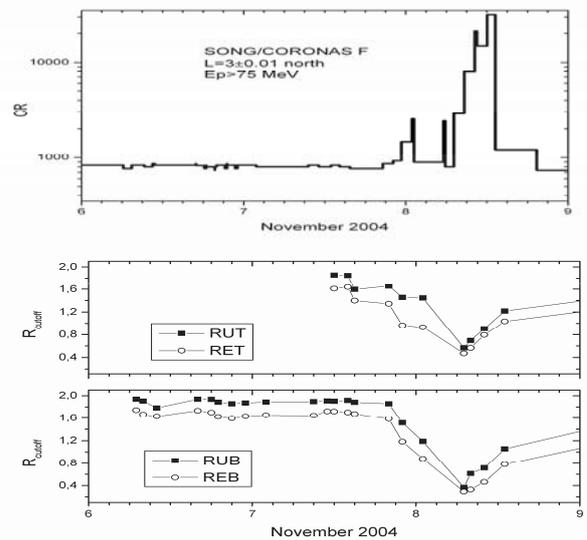

Fig.10. *The count rate of the channel 1 SONG at CORONAS-F at the crossings of L=3 (upper panel) and computed vertical cutoffs for the two models of geomagnetic field (T – Ts04, B – Ts89+Dst).*

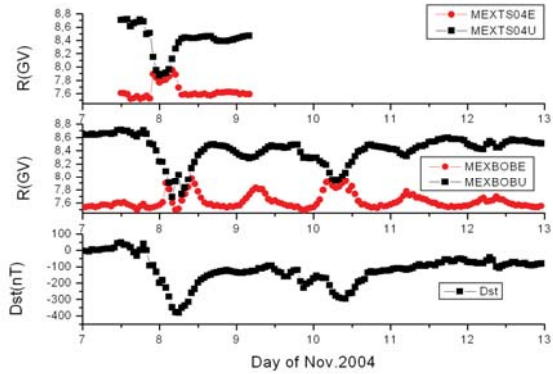

*Fig.11. From top to bottom: upper-U and effective-E vertical cutoff rigidities by Ts89+Dst (MEXBOB) and Ts04 model (MEXTS04) calculated for NM Mexico; Dst.*

better matching the time profile of increase at middle and low latitude neutron monitors than the approach (2).

Penetration of solar protons on CORONAS-F orbit to L<3 during the solar proton event on October 28-29, 2003 was observed due to the improved magnetospheric transmissivity. The time profile of the intensity of the penetrating protons (>75 MeV) approximately corresponds to that of effective and upper cut-off rigidity deduced from the approach (2). Ts89 model itself (without Dst extension) does not explain the observed profile.

During the geomagnetic disturbance on November 8, 2004, accompanied by galactic cosmic ray decrease observed both on neutron monitors and on high latitude passes of CORONAS-F below the atmospheric cut-off, the increase of >75 MeV protons on CORONAS-F due to the improved magnetospheric transmissivity was observed at L=3. For that case the differences between geomagnetic cut-offs computed using models 2 and 3 and their temporal evolutions are not significant. None of the models used match the time profile of energetic protons at L=3 precisely.

In summary, we have demonstrated that for mid-latitudes (at ground and at low altitudes) during strong geomagnetic disturbances the used models of external magnetic fields sometimes (for specific magnetospheric conditions) differ significantly in predictions. In addition, for one low latitude station was shown a difficulty when characterize geomagnetic transparency by concept of a single cutoff rigidity.

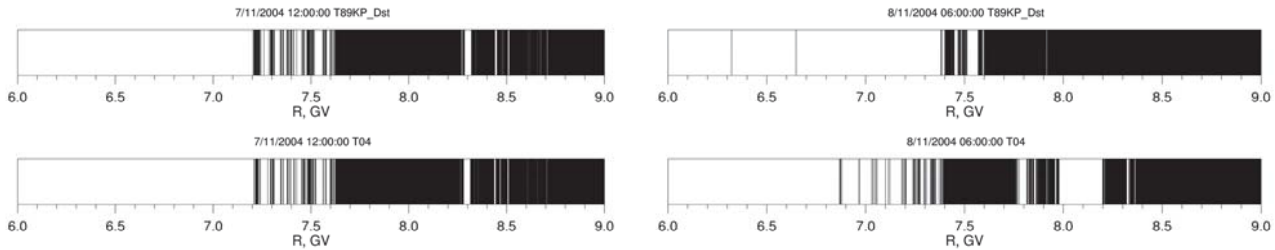

*Fig.12. The spectra of the allowed rigidities (black) in vertical direction for Mexcico NM station before the onset of the storm (Nov. 7, 2004, 12 UT, left) and during the Dst minimum (Nov. 8, 2004, 06 UT, right) for Ts89+Dst (top) and Ts04 (bottom) models.*

Comparison of cutoff rigidity profiles by two approaches (Ts89+Dst, Ts04) for position of Mexico NM is given in Fig. 11. Calculations by two models indicate that upper and effective cutoff rigidities vary in opposite manner; this is most pronounced during the minimum Dst. This illustrates that concept of a single cutoff value (effective or upper) is not good for strong disturbances. Fig. 12 demonstrates a complicated penumbra structure at Mexico NM station when magnetosphere is strongly disturbed. One can see the fine structures during the disturbance differ for used external magnetic field models.

## Conclusion

Vertical cut-off rigidities have been computed by using Tsyganenko geomagnetic field models (1) Ts89, (2) Ts89+ Dst extension and (3) Ts04 dynamical model depending on the availability of data, for selected intervals with different geomagnetic activity level in 2003 and 2004 while compared with NM and low altitude polar orbiting satellite measurements by CORONAS-F.

For November 20, 2003 strong disturbance the approach (3) provides more moderate shifts of vertical cutoffs and it is

## Acknowledgements

The work is supported by VEGA grant 2/4064. PIs and institutions of the NM and satellite data used are acknowledged: Rome (M.Storini), Athens (E.Mavromichala-ki), Oulu (I.Usoskin), Mexico (J.F.Valdes-Galicia), Haleakala (The University of Chicago, "National Science Foundation Grant ATM-9912341") and Moscow University (S.N.Kuznet-sov). Kp, Dst indices, IMF and solar wind data were obtained through the NSSDC OMNIWeb database. We thank the PIs of the MAG (N.F.Ness) and SWEPAM (D.J.McComas) ACE instruments due to preliminary IMF and verified solar wind data for November 2004.

## REFERENCES

[1] M.A.Shea, D.F.Smart, K.G.McCracken. "A Study of Vertical Cutoff Rigidity Using Sixth Degree Simulations of the Geomagnetic Field", *J. Geophys. Res.*, 70, 1965, pp4117-4130
[2] H.Miyasaka, K.Kudela, S.Shimoda et al. "Geomagnetic Cutoff Variation Observed with TIBET Neutron Monitor", *Proc. 28th ICRC*, Tsukuba, JP, Universal Academy Press, Inc. ed. T. Kajita et al., vol.6, 2003, pp3609-3612
[3] M.I.Tyasto, O.A.Danilova, N.G.Ptitsyna et al. "Cosmic Ray Cutoff Rigidities during Geomagnetic Storms: A Comparison of Magnetospheric Models", *Geomagn. Aeronomy*, 44, 2004, pp270-276

[4] E.O.Flückiger, R.Bütikofer, M.R.Moser, L.Desorgher. "The Cosmic Ray Ground Level Enhancement during the Forbush Decrease in January 2005", *Proc. 29th ICRC*, August 3-10, 2005, Pune, India
[5] D.F.Smart, M.A.Shea. "A comparison of the Tsyganenko Model Predicted and Measured Geomagnetic Cutoff Latitudes", *Adv. Space Res.*, 28, 2001, pp1733-1738
[6] R.A.Leske, R.A.Mewaldt, E.C.Stone, T.T.von Rosenvinge. "Observations of Geomagnetic Cutoff Variations during Solar Energetic Particle Events and Implications for the Radiation Environment at the Space Station", *J. Geophys. Res.*, 106, 2001, pp30011-30022
[7] D.F.Smart, M.A.Shea, M.J.Golightly, M.Weyland, A.S.Johnson. "Evaluation of the Dynamic Cutoff Rigidity Model Using Dosimetry Data from the STS-28 Flight", *Adv. Space Res.*, 31, 2003, pp841-846
[8] E.O.Flückiger, D.F.Smart, M.A.Shea. "A Procedure for Estimating the Changes in Cosmic Ray Cutoff Rigidities and Asymptotic Directions at Low and Middle Latitudes", *J. Geophys. Res.*, 91, 1986, pp7925-7930
[9] K.Kudela, M.Storini, M.Y.Hofer, A.Belov. "Cosmic Rays in Relation to Space Weather", *Space Sci. Rev.*, 93 (1-2), 2000, pp153-174
[10] J.W.Bieber, P.Evenson. "CME Geometry in Relation to Cosmic Ray Anisotropy", *Geophys. Res. Lett.*, 25 (15), 1998, pp2955-2958
[11] K.Munakata, J.W.Bieber, S.Yasue et al. "Precursors of Geomagnetic Storms Observed by the Muon Detector Network", *J. Geophys. Res.*, 105 (A12), 2000, pp27457-27468
[12] K.Leerungnavarat, D.Ruffolo, J.W.Bieber. "Loss Cone Precursors to Forbush Decreases and Advance Warning of Space Weather Effects", *Astrophys. J.*, 593 (1), 2003, pp587-596
[13] S.A.Starodubtsev, A.A.Turpanov, K.Kudela et al. "Real-time Cosmic Ray Distributed (RECORD) Database: A Status Report". *Proc. 29th ICRC*, August 3-10, 2005, Pune, India
[14] A.Chilingarian, V.Babayan, N.Bostanjyan, G.Karapetyan. "Correlated Measurements of the Secondary Cosmic Ray Fluxes by the Neutron Monitors and Muon Telescopes", *Int. J. Mod. Phys. A*, 20 (29), 2005, pp.6642-6645
[15] A.Chilingarian, V.Babayan, N.Bostonyan. "Aragats Space-environmental Centre: Status and SEP Forecasting Possibilities", *J. Phys. G*, 29 (5), 2005, pp.939-951
[16] H.Mavromichalaki, G.Souvatzoglou, C.Sarlanis et al. "The New Athens Center on Data Processing from Neutron Monitor Network in Real Time", *Ann. Geophys.*, 23, 2005, pp.3103-3110
[17] F.Jansen, *SEE-2005, International Symposium*, Nor Amberd, Armenia, 26-30 September 2005
[18] S.P.Kavlakov. "Global Cosmic Ray Intensity Changes, Solar Activity Variations and Geomagnetic Disturbances as North Atlantic Hurricane Precursors", *Int. J. Mod. Phys. A*, 20 (29), 2005, pp.6699-6701
[19] M.Storini. "Geomagnetic Storm: Effects on the Earths Ozone Layer", *Adv. Space Res.*, 27 (12), 2001, pp.1965-1974
[20] F.Spurny, K.Kudela, T.Dachev. "Airplane Radiation Dose Decrease during a Strong Forbush Decrease", *Space Weather*, 2 (5), 2004, Art. No. S05001
[21] I.L.Getley, M.L.Duldig, D.F.Smart, M.A.Shea. "Radiation Dose Along North American Transcontinental Flight Paths during Quiescent and Disturbed Geomagnetic Conditions", *Space Weather*, 2005, 3 (1): Art. No. S01004
[22] S.N.Kuznetsov, K.Kudela, S.P.Ryumin, Y.V.Gotselyuk. "CORONAS-F Satellite: Tasks for Study of Particle Acceleration", *Adv. Space. Res.*, 30, 2002, pp1857-1863
[23] S.N.Kuznetsov, I.N.Myagkova, S.P.Ryumin, K.Kudela, R.Bucik, H.Mavromichalaki. "Effects of the April 1994 Forbush Events on the Fluxes of the Energetic Charged Particles Measured on Board CORONAS-I: Their Connection with Conditions in the Interplanetary Medium", *J. Atmos. Sol-Terr. Phys.*, 64, 2002, pp535-539
[24] S.N.Kuznetsov, I.N.Myagkova, B.Y.Yushkov, Dynamics of the Boundary of Solar Electron Penetration into the Earth's Magnetosphere in November 2001, *Geomagn. Aeron.*, 45 (2), 2005, pp151-155
[25] N.A.Tsyganenko, M.I.Sitnov. "Modeling the Dynamics of the Inner Magnetosphere during Strong Geomagnetic Storms", *J. Geophys. Res.*, 110, 2005, A03208, doi:10.1029/2004JA010798
[26] J.Kassovicova, K.Kudela. *Preprint IEP SAS*, Kosice, Slovakia, 1995, pp45
[27] P.Bobik. *PhD dissertation*, P.J.Safarik U., Kosice, Slovakia, 2002
[28] K.Kudela, I.G.Usoskin. "On Magnetospheric Transmissivity of Cosmic Rays", *Czech. J. Phys.*, 54, 2004, pp239-254
[29] D.J.Cooke, J.E.Humble, M.A.Shea et al. "On cosmic-ray cutoff terminology", *Il. Nuovo Cimento C*, 14, 1991, pp213-234
[30] N.A.Tsyganenko. "A Magnetospheric Magnetic Field Model with a Warped Tail Current Sheet", *Planet. Space Sci.*, 37, 1989, pp5-20
[31] P.R.Boberg, A.J.Tylka, J.H.Adams Jr., E.O.Flückiger, E.Kobel. "Geomagnetic Transmission of Solar Energetic Protons during the Geomagnetic Disturbances of October 1989", *Geophys. Res. Lett.*, 22, 1995, pp1133-1136
[32] K.Kudela, M. Storini. "Direct and Indirect Relations of Cosmic Rays to Space Weather", *ESA SP-477*, 2002, pp289-292